\documentclass[useAMS,usenatbib]{mn2e}

\usepackage{graphicx,amssymb,amsmath}

\title[Spectroscopy of Ter 8]{Spectroscopy of Red
Giants in the globular cluster Terzan 8: kinematics and evidence for the surrounding
Sagittarius stream}
\author[Sollima et al.]{A. Sollima$^{1}$\thanks{E-mail:
antonio.sollima@oabo.inaf.it}, E. Carretta$^{1}$, V. D'Orazi$^{2,3}$, R. G.
Gratton$^{4}$, A. Bragaglia$^{1}$, S. Lucatello$^{4}$\\
$^{1}$ INAF Osservatorio Astronomico di Bologna, via Ranzani 1, Bologna, 40127,
Italy\\
$^{2}$ Department of Physics and Astronomy, Macquarie University, Balaclava
Road, North Ryde, NSW 2109, Australia\\
$^{3}$ Monash Centre for Astrophysics, School of Mathematical Sciences, Monash
University, Building 28, Clayton, VIC 3800, Australia\\
$^{4}$ INAF Osservatorio Astronomico di Padova, vicolo dell'Osservatorio 5,
Padova, 35122, Italy}
\begin{document}

%\date{Accepted 2013 May 10. Received 2013 May 10; in original form
%2013 May 10}

\pagerange{\pageref{firstpage}--\pageref{lastpage}} \pubyear{2014}

\maketitle

\label{firstpage}

\begin{abstract}
We present the results of a spectroscopic survey of Red Giants in the globular
cluster Terzan 8 with the aim of studying its kinematics. We derived accurate
radial velocities for 82 stars located in the innermost 7$\arcmin$ from the
cluster center identifying 48 bona fide cluster members. The kinematics
of the cluster have been compared with a set of dynamical models accounting for
the effect of mass segregation and a variable fraction of binaries. The derived
velocity dispersion appears to be larger than that predicted for mass-segregated 
stellar systems without binaries, indicating that either the cluster is 
dynamically young or it contains a large fraction of binaries ($>$30\%). 
We detected 7 stars with a radial velocity compatible with the cluster systemic 
velocity but with chemical patterns which stray from those of both the cluster 
and the Galactic field. These stars are likely members of the Sagittarius 
stream surrounding this stellar system.
\end{abstract}


\section{Introduction}
\label{intro_sec}

\begin{figure*}
 \includegraphics[width=13cm]{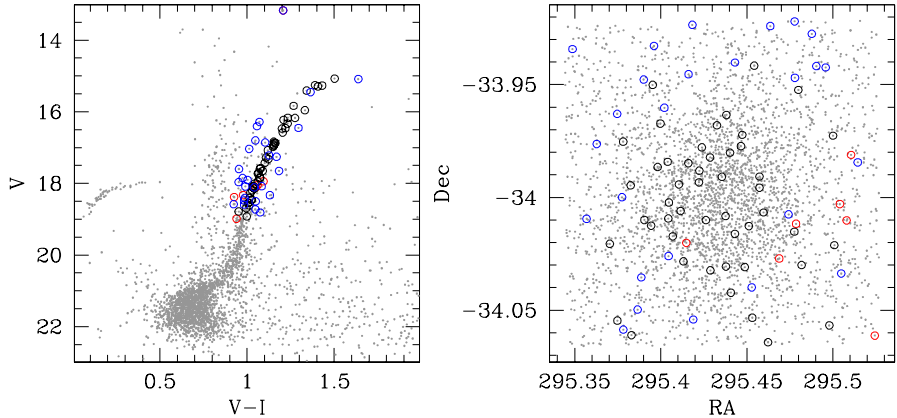}
 \caption{V,V-I color-magnitude diagram (left panel) and map (right panel) of 
 Terzan 8 (grey points; from 
 Montegriffo et al. 1998). In both panels black, blue and red dots (asterisks,
 open and filled dots in the printed version of the paper) represent 
 the observed cluster members, 
 field stars and Sagittarius stars, respectively.}
\label{cmd}
\end{figure*}
 
The globular cluster (GC) Terzan 8 is one of the faintest GCs populating the
outer halo. Discovered by Terzan (1968), it is located at a distance of 19.4 kpc
from the Galactic center in the constellation of Sagittarius ($RA=19^{h}41^{m}44.41^{s},~
Dec=-33^{\circ}59^{\arcmin}58.1^{\arcsec}$; Harris 1996, 2010 edition). The 
photometric studies of this GC (Ortolani \& Gratton 1990; Montegriffo et al.
1998) suggested that it is
an old, metal-poor object with a blue Horizontal Branch (HB). 
The first spectroscopic study of this cluster analysed 3 Red Giant stars in the
CaII triple region and derived a very low metallicity ([Fe/H]=-2.3; Da Costa \& 
Armandroff 1985). More recently, Mottini, Wallerstein \& McWilliam (2008) used
high-resolution spectroscopy to determine abundances of Fe, O,
Na, $\alpha$-, Fe-peak, and neutron-capture elements for 3 Red Giants. 
The small number of targets of the above studies did not allow a relevant
analysis of the cluster kinematics.
Recently, we performed an high-resolution spectroscopic survey
of this cluster (Carretta et al. 2014) providing accurate radial velocities for
82 stars in the central region of the cluster and abundances of 16 members. 

From the dynamical
point of view, Terzan 8 is the less concentrated GC of the Milky Way (c=0.41)
and it appears very extended (with an half-light radius of 15.3 pc; Salinas et 
al. 2011). Such a peculiar structure implies a half-mass relaxation time 
comparable with its age, suggesting that Terzan 8 could be dynamically young. 

The interest for this cluster increased because of its probable association with the
Sagittarius dwarf galaxy (Ibata, Gilmore \& Irwin 1994). Indeed, the location in the sky,
radial velocity and distance of Terzan 8 are consistent with that of this
satellite galaxy (Da Costa \& Armandroff 1995; Dinescu et al. 2001; 
Palma, Majewski \& Johnston 2002; Bellazzini et al. 2003; Majewski et al. 2004; Carraro, Zinn
\& Moni Bidin 2007). In particular, N-body 
simulations indicate that Terzan 8 could be immersed in the trailing arm of
the Sagittarius stream (Edelsohn \& Elmegreen 1997; Law \& Majewski 2010).
Moreover, the presence of the Sagittarius stream has been detected as an
overdensity of stars in a wide region surrounding this cluster 
(Giuffrida et al. 2010; Siegel et al. 2011).

In this paper we use the radial velocities obtained in the survey by Carretta et
al. (2014) to investigate the kinematics of Terzan 8. In Sect. 2 we describe the
observational material and the data reduction procedure. Sect. 3 is devoted to 
the comparison of the derived velocity dispersion with a set of 
dynamical models. In Sect. 4 we discuss the spectroscopic signature of the
surrounding Sagittarius stream. We summarize and discuss our results in Sect. 5.

\section{Observations and Data reduction}
\label{obs_sec}

\begin{figure}
 \includegraphics[width=8.7cm]{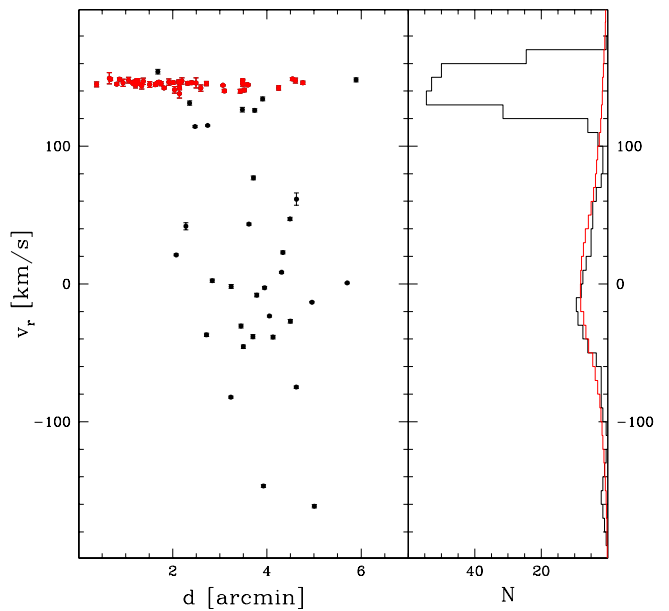}
 \caption{Radial velocities of the target stars as a function of the distance
 from the cluster center (left panel). Red points (filled dots in the printed
 version of the paper) mark the bona fide cluster
 members (see Sect. \ref{kin_sec}). In the right panel the distribution of
 radial velocities is shown. The prediction of the Galactic model of Robin et
 al. (2003) is shown with red histograms (grey in the printed version of the
 paper).}
\label{rv}
\end{figure}

Spectra have been obtained using the multi-object spectrograph VLT/FLAMES 
(Pasquini et al. 2002). A more detailed description of the observations and analysis is presented in
Carretta et al. (2014). We give here only a short description.
We used the Ultraviolet and Visual Echelle Spectrograph (UVES) 580 nm setup 
($4800-6800 \AA$; with a spectral resolution of $R\sim45000$) and the 
GIRAFFE high-resolution setups 
HR11 and HR13, providing a
spectral resolution of R$\sim$24200 and 22500, respectively.
The target stars have been selected from the photometric catalog by Montegriffo
et al. (1998) covering a region of $9.15\arcmin\times8.6\arcmin$ centered on the
cluster. We selected seven among the brightest Red Giant Branch (RGB) stars for 
the UVES fibres ($R\sim45000$) while the GIRAFFE fibres 
were allocated to objects on or near the RGB (84 stars) or the HB (17
stars; see Fig. \ref{cmd}). 
A set of $6\times4720$ sec exposures have been obtained. Unfortunately, 
because of the instrumental setup, the low metallicity and S/N, we could not 
measure radial velocities for 2 RGB and all HB stars. The observations were obtained in service mode under
seeing conditions of $0.7\arcsec<FWHM<2.2\arcsec$. 
The spectra were reduced (bias and flat-field corrected, 1D extracted, and 
wavelength calibrated) using the ESO pipeline. We applied sky subtraction and 
division by an observed early-type star (UVES), or a synthetic spectrum 
(GIRAFFE) to correct for telluric features near the [OI] line using the IRAF 
routine telluric. The latter correction was applied only to the UVES and bright 
GIRAFFE samples. We shifted all spectra according to their heliocentric velocity 
and combined the individual exposures. The UVES final spectra have 
S/N in the range 45-80 while the GIRAFFE spectra have S/N values ranging from 5 
to 80. 
The heliocentric radial velocities
and their uncertainties were then measured on the combined spectra of 82 stars 
using the IRAF task rvidlines. Uncertainties have been calculated from
the r.m.s. of the velocities measured using different lines. They
are listed in Table 2 of Carretta et al. (2014). 

In Fig. \ref{rv} the radial velocities of the observed stars are plotted as a
function of the distance from the cluster center. The signal of Terzan 8 is
clearly visible as a group of stars with velocity $v_{r}\sim142$ km/s and a
small ($<2$ km/s) dispersion. The Galactic field population is also clearly
identifiable at velocities $v_{r}<100$ km/s. We note that at $v_{r}>100$ km/s, 
besides the group of cluster members, other 8 stars have velocities similar to
that of Terzan 8 but with a larger spread. Seven of them are located in the 
color-magnitude diagram (CMD) close to the cluster RGB, while one of them is
clearly a bright field star at $V<14$ (see Fig. \ref{cmd}). 
Curiously, all these 7 stars are located in the same half of the observed field 
of view. All these stars were
observed with GIRAFFE.

For 5 out these 7 stars we derived the abundance of Fe (from both FeI and
FeII lines), Mg, Si, Ca, Ti, Ba from
their spectra. The other two stars have a faint magnitude ($V>18.5$) and the
low S/N ($\sim20$) of their spectra do not allow a reliable estimation of their abundances.
We performed the analysis as described in Carretta et al. (2014).
Temperatures have been derived using the V-I color-temperature relation by
Alonso, Arribas \& Mart{\'{\i}}nez-Roger (1999) and the reddening E(B-V)=0.12 (Schlegel, Finkbeiner \& 
Davis 1998; Harris catalog). Gravities have been derived using the universal gravitation law 
assuming a mass of 0.85 $M_{\odot}$ (suitable for a Red Giant) and a luminosity
calculated converting the V band into bolometric magnitude using the corrections
by Alonso et al. (1999), and assuming the apparent 
distance modulus of Terzan 8 $(m-M)_{V}=17.47$ (from the Harris
catalog), the above mentioned reddening and a solar bolometric magnitude of
$M_{bol,\odot}=4.75$. Microturbulence velocities have been calculated as a
function of gravities using the relation by Worley et al. (2013).
We interpolated within 
the Kurucz (1993) grid of model atmospheres (with the option for overshooting 
turned on) and tuned the abundances of the various elements to match the 
equivalent widths of several lines.

\section{Kinematics}
\label{kin_sec}

\subsection{Overall velocity dispersion, density profile and mass function}
\label{met_sec}

\begin{figure}
 \includegraphics[width=8.7cm]{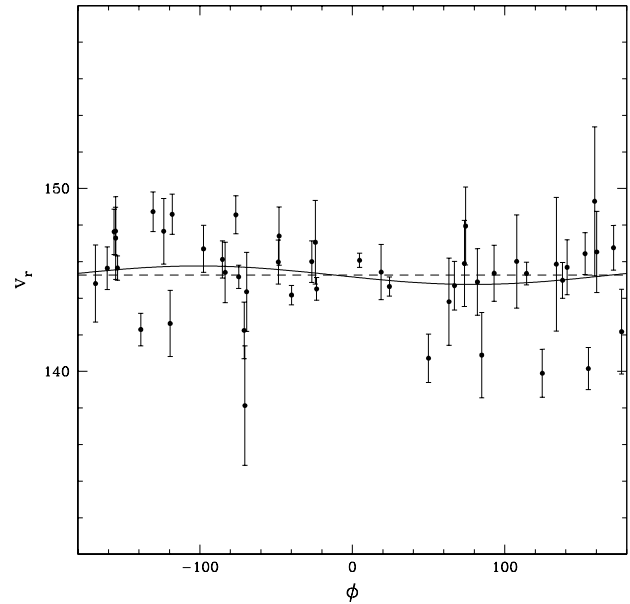}
 \caption{Radial velocity of the 48 member stars of Terzan 8 as a function of
 the position angle. The best-fit sinusoidal trend and the mean radial velocity 
 are marked with solid and dashed lines, respectively.}
\label{rot}
\end{figure}

\begin{figure}
 \includegraphics[width=8.7cm]{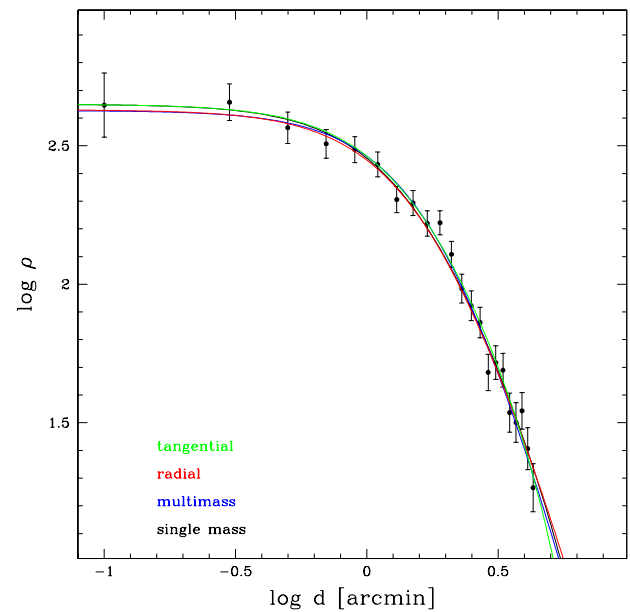}
 \caption{Density profile of Terzan 8. The prediction of the various models are
 overplotted.}
\label{dens}
\end{figure}

\begin{figure}
 \includegraphics[width=8.7cm]{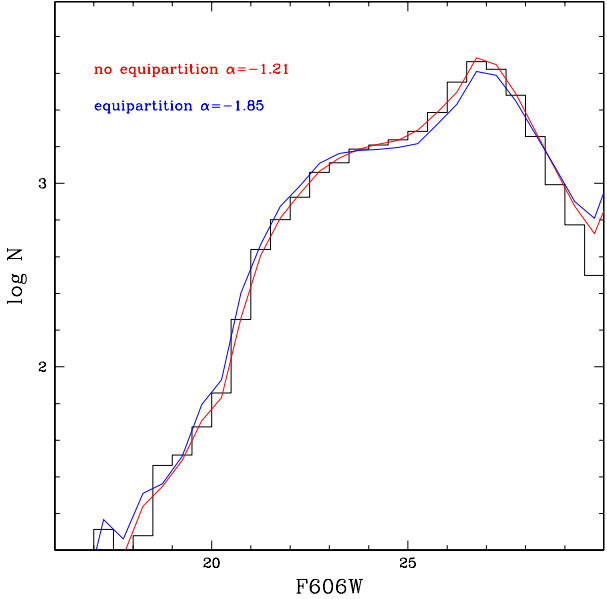}
 \caption{F606W luminosity function of Terzan 8 (black histograms). The
 prediction of models with and without kinetic energy equipartition are
 also marked with red and blue lines (black and grey in the printed version of
 the paper), respectively.}
\label{lf}
\end{figure}

To study the structure and kinematics of Terzan 8 we derived its overall
velocity dispersion, its density profile and its luminosity function (MF).
As a first step, we selected a sample of bona fide cluster
members on the basis of their locations in the CMD and their radial
velocities.
In particular, we considered only those stars with
\begin{itemize}
\item{a location in the CMD within $\Delta(V-I)<0.1$ around the cluster RGB mean
ridge line and a magnitude within $14<V<19$;}
\item{a radial velocity within 5$\sigma$ from the
mean systemic velocity.}
\end{itemize}

According to the above cirteria we defined a sample of 48 members. The average
radial velocity turns out to be $\langle v_{r}\rangle=145.26\pm0.15$ km/s
(where the reported uncertainty is the error on the mean, while r.m.s=2.46 km/s).
For comparison, Da Costa \& Armandroff (1995) reported an average radial velocity $\langle
v_{r}\rangle=130\pm8$ from the analysis of three stars. The star-to-star
comparison with our catalog gives a perfect agreement and the difference in the
average velocity is probably due to their very small sample. 
We cannot compare our values with the study of Mottini et al. (2008) because they do not 
provide the radial velocities for their targets. 

The overall velocity dispersion has been calculated using the  
algorithm proposed by Pryor \& Meylan (1993), where the velocity dispersion is
calculated as the quantity $\sigma_{v}$ that maximizes the likelihood
\begin{eqnarray}
l&=&\sum_{i} ln \int_{-\infty}^{+\infty}
\frac{exp\left[-\frac{(v'-\bar{v})^{2}}{2\sigma_{v}^{2}}-\frac{(v_{i}-v')^{2}}{2\delta_{i}^{2}}\right]}{2\pi\sigma_{v}\delta_{i}^{2}}
dv'\nonumber\\
&=&-\frac{1}{2}\sum_{i}\left(\frac{(v_{i}-\bar{v})^{2}}{\sigma_{v}^{2}+\delta_{i}^{2}}+ln[2\pi(\sigma_{v}^{2}+\delta_{i}^{2})]\right)
\label{sig_eq}
\end{eqnarray}
where $v_{i}$ and $\delta_{i}$ are the velocity of the i-th star and its associated 
uncertainty. The calculated velocity dispersion turns out to be
$\sigma_{v}=1.72\pm0.14$ km/s.
Unfortunately, because of the small number of target stars it is not possible 
to construct a reliable velocity dispersion profile.
In Fig. \ref{rot} the radial velocities of the bona fide cluster members are
plotted as a function of the position angle. The distribution appears
homogeneous and no clear trend is apparent. A fit with a sinusoidal curve
indicates a maximum rotation amplitude of $v_{rot,max}~sin i=0.50\pm0.21$ km/s 
consistent with no significant rotation along the line-of-sight within Terzan 8. 

The density profile has been constructed from the photometric catalog by 
Montegriffo et al. (1998) by dividing the number of stars brighter than $V<21$
and contained in circular annuli by the annulus area. We adopted the above magnitude 
limit to ensure a significant number of stars with a 
completness level $>$90\%, as estimated through artificial star experiments.
The derived density profile is shown in Fig. \ref{dens}.

The luminosity function of the cluster has been derived using the deep ACS@HST 
photometry provided by the ACS treasury project (Sarajedini et al. 2007).
These data cover a region of $202\arcsec\times202\arcsec$ around Terzan 8 and
provide a deep CMD in the F606W and F814W filters down to $F666W\sim28$. 
A detailed description of the 
photometric reduction, astrometry, and artificial 
star experiments for this database can be found in Anderson et al. (2008).
We selected Main Sequence stars by selecting stars
within 3 times the standard deviation of the median F666W-F814W color at a 
given F606W magnitude. The F606W luminosity function has been then calculated by
counting stars in bins of magnitudes of 0.5 mag width and it is shown in Fig. \ref{lf}.  

\subsection{Models}

\begin{table*}
 \centering
 \begin{minipage}{140mm}
  \caption{Properties of the adopted dynamical models.}
  \begin{tabular}{@{}lcccccccccr@{}}
  \hline
model & equipartition & $r_{a}$ & $W_{0}$ & $r_{h}$   & $\alpha$ & $\xi$ & log $M/M_{\odot}$ &
$\langle\sigma_{v}\rangle$ & $t_{rh}$ & $(M/L_{V})_{dyn}$\\
      &               & pc      &         & pc        &          &   \%  &                   &
km/s                       & Gyr      & $M_{\odot}/L_{\odot}$\\
 \hline
s\_iso\_fb0  & no  & $\infty$ & 5.0 & 24.36 & -1.21 & 0  & 4.95 & 1.49  & 1.99 & 1.2\\
s\_iso\_fb20 & no  & $\infty$ & 5.0 & 24.36 & -1.21 & 20 & 4.95 & 1.61  & 1.99 & 1.1\\
s\_iso\_fb40 & no  & $\infty$ & 5.0 & 24.36 & -1.21 & 40 & 4.95 & 1.75  & 1.99 & 0.9\\
s\_rad\_fb0  & no  & 23.36    & 4.0 & 26.72 & -1.21 & 0  & 5.00 & 1.52  & 2.38 & 1.2\\
s\_tan\_fb0  & no  & 36.00    & 5.0 & 22.68 & -1.21 & 0  & 4.93 & 1.48  & 1.75 & 1.3\\
m\_iso\_fb0  & yes & $\infty$ & 9.0 & 43.44 & -1.85 & 0  & 4.97 & 1.06  & 7.47 & 7.4\\
m\_iso\_fb20 & yes & $\infty$ & 9.0 & 43.44 & -1.85 & 20 & 4.97 & 1.24  & 7.47 & 5.4\\
m\_iso\_fb40 & yes & $\infty$ & 9.0 & 43.44 & -1.85 & 40 & 4.97 & 1.43  & 7.47 & 4.1\\
\hline
\end{tabular}
\end{minipage}
\end{table*}

\begin{figure*}
 \includegraphics[width=13cm]{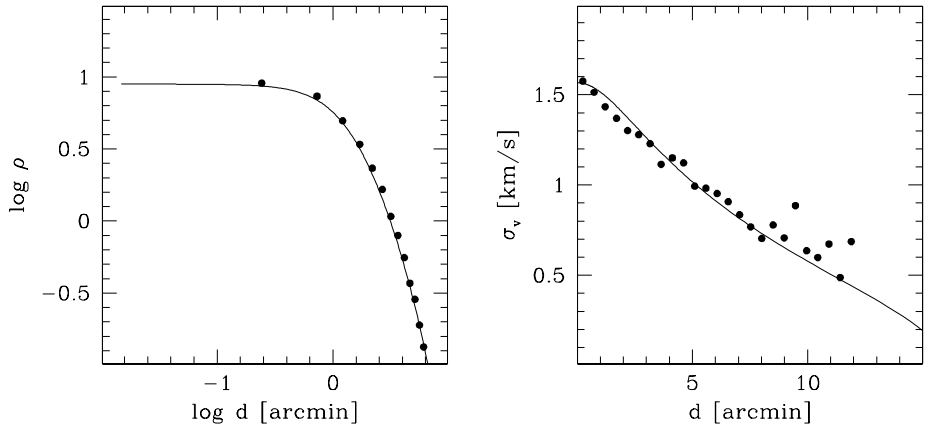}
 \caption{Projected density (left panel) and velocity dispersion (right panel) 
 profiles calculated from the last snapshot of the N-body simulation. The 
 quantities at the end of the simulation are marked by filled dots. 
 The best-fit static King (1966) model is overplotted in both panels with 
 solid lines.}
\label{nbody}
\end{figure*}

We compared the overall velocity dispersion of Terzan 8 as measured in
our sample with the prediction of a set of dynamical models taking into account
the location of targets across the field of view. 

In principle, in relaxed stellar systems the most massive RGB stars are
kinematically colder than low-mass stars. Therefore, since our velocity
dispersion is calculated only for RGB stars, we should take into
account this effect.
However, given the long relaxation time of Terzan 8 it is possible that the cluster has
still not reached a significant degree of kinetic energy equipartition. In this
case, stars with different masses would follow the same velocity dispersion
profile.

Therefore, we considered a set of single-mass King (1966) and multi-mass 
King-Michie models (Gunn \& Griffin 1979). For single-mass models we also
considered two extreme degrees of radial and tangential anisotropy using the
formalism by Gunn \& Griffin (1979; we assumed
the criterion for stability by Nipoti, Londrillo \& Ciotti 2002).
For multi-mass models we considered 8 mass bins between
0.1 and 0.9 $M_{\odot}$ having a width of 0.1$M_{\odot}$ each. 

Dark remnants have been added
using the initial-final mass relation and the retention fractions by Kruijssen 
(2009). We adopted for the dark remnants the same MF of the other
stars. 

A population of binaries has been also added by calculating the projected
velocity component of a 0.85 $M_{\odot}$ primary star in a binary systems with a
given period, eccentricity and inclination angles (McConnachie \& C{\^o}t{\'e} 2010). 
For this purpose, the
secondary masses have been randomly extracted from a uniform distribution
between $0.1<M/M_{\odot}<0.85$. A period has been then extracted from the log-normal distribution
of Duquennoy \& Mayor (1991) truncated at the periods corresponding to
the limiting semi-major axes $a_{min}$ and $a_{max}$. The value of $a_{min}$ has
been chosen as the minimum distance at which mass-transfer between the 
components would occur (Lee \& Nelson 1988) while for $a_{max}$ has been chosen
the value for which the binding energy of the binary equates the mean kinetic
energy of clusters stars ($\langle m\sigma_{v}^{2}\rangle$). The eccentricity
has been chosen following the prescription of Duquennoy \& Mayor (1991).
We adopted a random distribution of orbital phases, inclination angles and 
longitudes of the periastron.

For each model, we searched for the parameters (central adimensional potential
$W_{0}$, MF slope $\alpha$, and core radius $r_{c}$) which bestfit the observed
density profile shown in Fig. \ref{dens}.
A synthetic CMD has been contructed by interpolating across the
set of tracks by Dotter et al. (2007) with appropriate metallicity and age and
adopting the distance modulus and reddening described in Sect. \ref{obs_sec}.
Stars have been distributed across the cluster according
to the density profile of their mass bin. The F606W luminosity function of the
synthetic CMD has been then estimated using the same technique described in
Sect. \ref{met_sec} after correcting for the completeness estimated through the
large set of artificial stars experiments described in Anderson et al. (2008). 
We considered only those stars within 
a distance of $200\arcsec$
from the cluster center. The slope of the MF has been then chosen to reproduce
the observed F606W luminosity function. We adopted the algorithm described in
Sollima, Bellazzini \& Lee (2012) to derive the bestfit parameters through an iterative
procedure. 

The cluster mass has been then calculated 
by matching the
number of stars brighter than $V<20.5$ in the innermost $5\arcmin$ of the
photometric catalog by Montegriffo et al. (1998). 

It is known that the interaction with the Milky Way tidal field can alter the 
velocity dispersion of the system by heating stars at large distances from 
the center (K{\"u}pper et al. 2010). To estimate the effect of tides we performed an
N-body simulation to follow the structural and kinematical evolution of a 
cluster with the orbital and structural characteristics of Terzan 8. We used 
the last version of the collisionless N-body code gyrfalcON (Dehnen 2000). 
The cluster has been represented by 10,000 particles distributed according to 
the best-fit King (1966) model. The mean masses of the particles have been 
scaled to match the cluster mass predicted by this model. We adopted a leapfrog 
scheme with a time step of $\Delta t = 5.82 \times 10^{4}$ yr and a softening 
length of 0.12 pc (following the prescription of Dehnen \& Read 2011). 
Such a relatively large time step and softening length do not affect the 
accuracy of the simulation because we are interested in the tidal effects that 
occur on a timescale significantly shorter than the relaxation time (the 
timescale at which collisions become important). The cluster was launched 
within the three-component (bulge + disk + halo) static Galactic potential of 
Johnston, Spergel \& Hernquist (1995) on a quasi-circular orbit starting from it present-day
location and its evolution has been 
followed for 1.2 Gyr (corresponding to two orbital periods). 
The projected density and velocity dispersion profiles at the end of the 
simulation are shown in Fig. \ref{nbody}. It is apparent that both the density
and the velocity dispersion profile of the cluster are almost unaffected by the
presence of the tidal field with only a hint of deviation from the prediction of
the static model at distances $d>8\arcmin$. It is also interesting to note that
in spite of the extended structure of the cluster, the tidal field does not
produce a significant distortion of the cluster structure within the field of
view covered by our photometric observations (with an ellipticity
$\epsilon<0.002$ within 4.5$\arcmin$ from the cluster center).

To calculate the overall velocity dispersion predicted by the various models we
adopted the Monte Carlo approach already applied in Sollima et al. (2012).
In particular, for each model the following steps have been performed:
\begin{itemize}
\item{We compute the distance of the star from the cluster center and extract a 
velocity from a Gaussian function with a standard deviation corresponding to 
the local line-of-sight (LOS) velocity dispersion in 
the model. When the N-body simulation has been considered, the projected velocity
of the closest particle has been adopted;}
\item{We extract a random velocity from a Gaussian distribution with dispersion
$\delta_{i}$ and sum it to the previous one to compute a simulated observational
velocity;}
\item{A random number between 0 and 1 is extracted from a uniform distribution. 
If this number is smaller than the adopted binary fraction ($\xi$) a binary is randomly 
extracted from the library (see above) and its apparent velocity is added to 
the simulated velocity.}
\end{itemize}
In this way, every realization contains the same number of objects at the same 
location as the observed sample. Finally, the velocity dispersion of the 
simulated sample is calculated using Eq. \ref{sig_eq}. The above procedure is 
repeated 1000 times to compute the distribution of predicted velocity 
dispersions. The outcome of all the simulations performed is summarized in 
Table 1.

\subsection{Results}

\begin{figure*}
 \includegraphics[width=13cm]{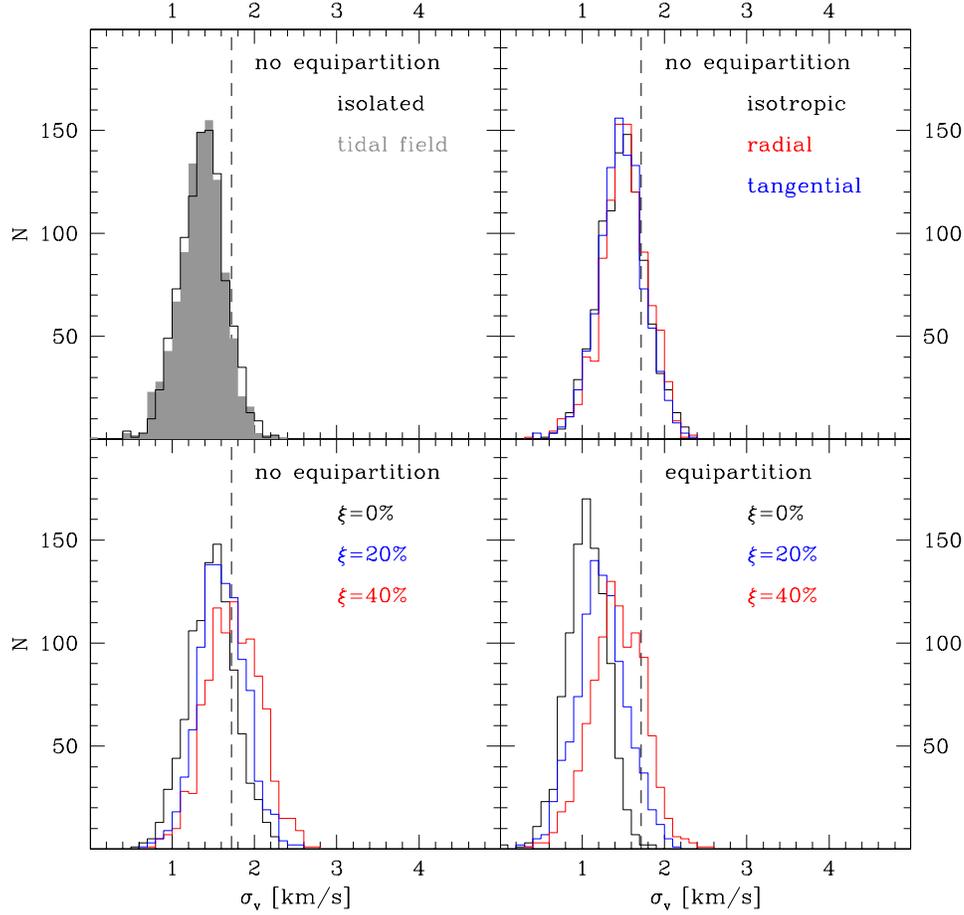}
 \caption{Predicted velocity dispersion of the considered dynamical models. In
 the bottom panels models without (bottom-right panel) and 
 with (bottom-left panel) kinetic energy equipartition and different fraction of
 binaries are compared. In the top-left panel the distributions of the 
 isolated model and 
 the N-body simulation within the tidal field are compared. 
 In the top-right panel two models with different degrees of anisotropy are
 compared. The observed velocity
 dispersion is marked with a dashed line in all panels.}
\label{mod}
\end{figure*}

The distribution of predicted LOS velocity dispersions for the considered models
are compared with the observed measure in Fig. \ref{mod} (bottom panels). It is apparent that
the model without binaries and accounting for the effect of mass segregation
predicts a velocity dispersions which is significantly smaller than the observed value of 
Terzan 8. The observed discrepancy can be overcome assuming a large fraction of 
binaries ($\xi>30\%$). It is also interesting to note that a very steep MF
has been derived in this case ($\alpha=-1.85\pm0.15$; for reference a Salpeter 1955 MF
has slope $\alpha=-2.35$ and the slopes derived for 17 GCs by the comprehensive 
study by Paust et al. 2010 range between -0.32 and -1.69).

A different situation is apparent when models without equipartition are 
considered: in this case models with a moderate fraction of binaries predict
velocity dispersion compatible within the errors with the observed value.
In this case, the derived MF slope turns out to be $\alpha=-1.21\pm0.09$.

There is no appreciable difference between the best-fit single-mass 
King (1966) model in isolation and the Nbody simulation within the Milky Way 
tidal field (see the top-left panel of Fig. \ref{mod}). This is expected since the small deviations from the prediction of
the isolated models occurs only at distance larger than the radial coverage of
the target stars.

In the same way, models with different degrees of anisotropy predict similar
velocity dispersions (see the top-right panel of Fig. \ref{mod}). This is due to the distribution of the target stars 
within the field of view: in the region where most of the observed stars reside,
models with different anisotropies predict very similar velocity dispersion
profiles.

\section{The Sagittarius stream around Terzan 8}

\begin{table*}
 \centering
 \begin{minipage}{140mm}
  \caption{Abundances of the Sgr stars.}
  \begin{tabular}{@{}lcccccccr@{}}
  \hline
ID & [Fe I/H] & [Fe II/H] & [Mg/Fe] & [Si/Fe] & [Ca/Fe] & [Ti I/Fe] &
  [$\alpha$/Fe] & [Ba/Fe]\\
 \hline
%36 & -1.34 & --    & --    & 0.43  & -0.55 & --    & -0.06 & -1.46\\
%   & ($\pm$0.43) & &       & ($\pm$0.24) & ($\pm$0.30) &   & ($\pm$0.69) & ($\pm$0.30)\\
41 & -1.44 & -1.43 &  0.27 &  0.40 &  0.29 & -0.05 &  0.23 & -0.18\\
   & ($\pm$0.26) & ($\pm$0.36) & ($\pm$0.30) & ($\pm$0.20) & ($\pm$0.17) & ($\pm$0.10) & ($\pm$0.19) & ($\pm$0.30)\\
47 & -0.83 & -0.86 & -0.17 &  0.17 &  0.33 & -0.02 &  0.08 & 0.43\\
   & ($\pm$0.34) & ($\pm$0.38) & ($\pm$0.10) & ($\pm$0.32) & ($\pm$0.28) & ($\pm$0.10) & ($\pm$0.22) & ($\pm$0.30)\\
57 & -0.84 & -0.85 & -0.08 &  0.29 &  0.30 & -0.08 & 0.11 & 0.36\\
   & ($\pm$0.35) & ($\pm$0.10) & ($\pm$0.30) & ($\pm$0.13) & ($\pm$0.16) & ($\pm$0.28) & ($\pm$0.22) & ($\pm$0.30)\\
65 & -1.08 & -1.16 & -0.36 & 0.29 & 0.11 & 0.04 & 0.02 & 0.27\\
   & ($\pm$0.35) & ($\pm$0.30) & ($\pm$0.30) & ($\pm$0.16) & ($\pm$0.29) & ($\pm$0.10) & ($\pm$0.28) & ($\pm$0.30)\\  
70 & -1.14 & -1.11 & 0.16 & 0.39 & 0.24 & 0.67 & 0.37 & 0.26\\
   & ($\pm$0.39) & ($\pm$0.31) & ($\pm$0.30) & ($\pm$0.43) & ($\pm$0.24) & ($\pm$0.18) & ($\pm$0.22) & ($\pm$0.30)\\
%94 & -1.91 & --  & 0.32 & 1.44 & 0.11 & 1.44 & 0.83 & -0.80\\ 
%   & ($\pm$0.47) & & ($\pm$0.30) & ($\pm$0.30) & ($\pm$0.63) & ($\pm$0.84) & ($\pm$0.71) & ($\pm$0.30)\\ 
\hline
\end{tabular}
\end{minipage}
\end{table*}

\begin{figure*}
 \includegraphics[width=13cm]{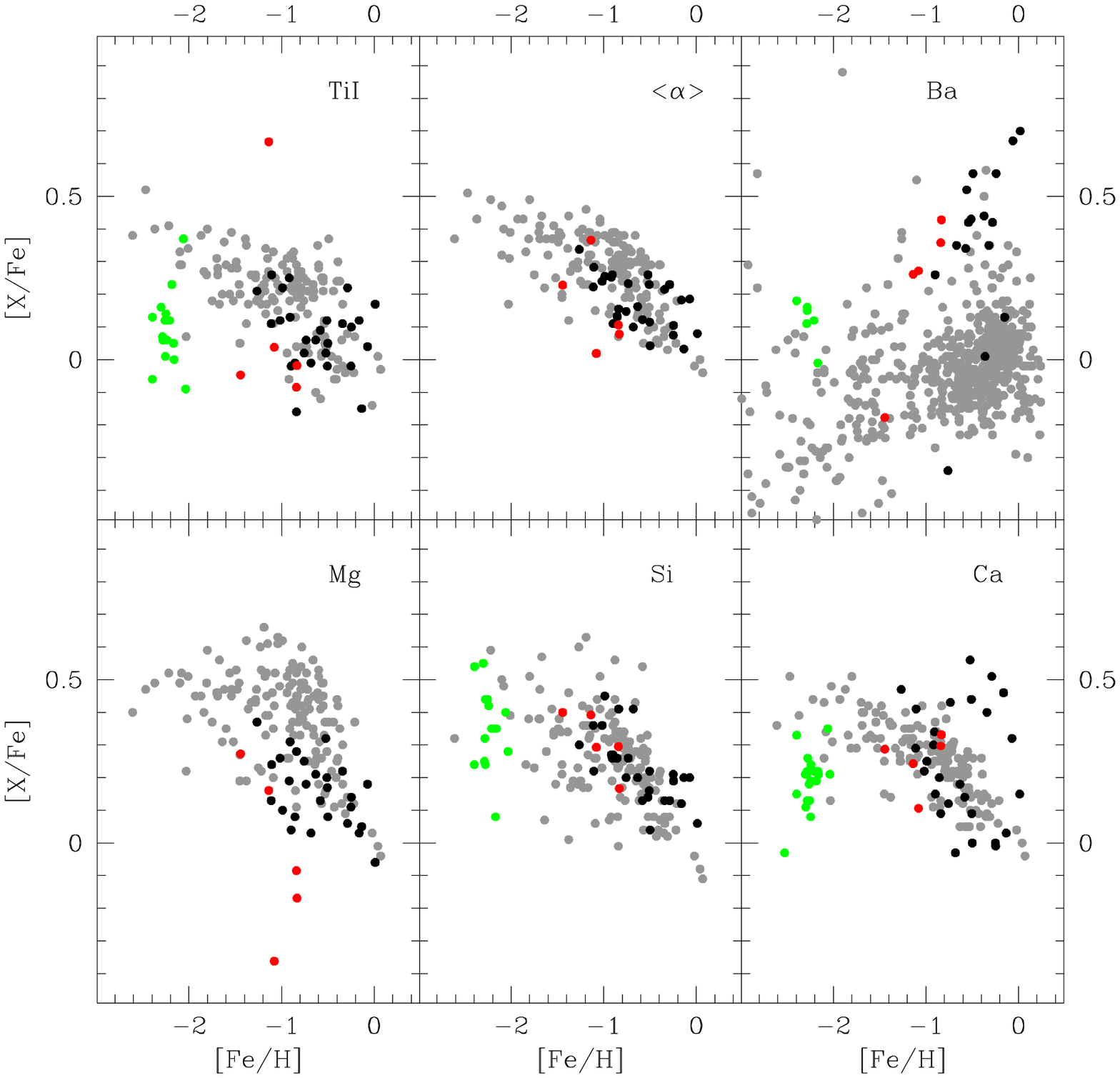}
 \caption{Abundance patterns of the $\alpha$-elements and Ba in the 5 stars suspected
 to belong to the Sagittarius stream (red dots; open stars in the printed
 version of the paper). The abundances of the field
 giants (grey dots; from Gratton et al. 2003 and Venn et al. 2004), Sagittarius stars (black dots;
 from Carretta et al. 2010 and Sbordone et al. 2007) and Terzan 8 stars (green dots, open dots in the
 printed version of the paper; from Carretta et al.
 2014) are also plotted.} 
% In the top-right panel the difference between the
% abundances of Fe I and Fe II are shown for the 7 stars suspected
% to belong to the Sagittarius stream.}
\label{alpha}
\end{figure*}

As already reported in Sect. \ref{obs_sec}, a group of 7 stars located close to
the RGB of Terzan 8 with a similar systemic velocity but a larger spread are
present. These stars could be either {\it i)} binary stars of Terzan 8, {\it
ii)} Galactic field interlopers, or {\it iii)} stars belonging to the
Sagittarius stream surrounding Terzan 8.

We compared the observed distribution of velocities with the Galactic model of
Robin et al. (2003). For this purpose, a synthetic catalog covering 1 sq.deg.
centered at the cluster location has been retrieved. We selected only stars
which lie in the CMD brighter than $V<19$ and within $\Delta(V-I)<0.1$ about the
RGB mean ridge line and normalized the sample to reproduce the
observed number of stars at $v_{r}<100$ km/s (where only the Galactic field
population is expected to be present). The model predicts a contamination of 
3.7 stars at $v_{r}>100$ km/s. Note that all the synthetic field stars in this
velocity range are Main Sequence stars in the foreground of the cluster.
Unfortunately, the time coverage of our spectra does
not allow to identify binaries from their velocity variation, so the only chance
to distinguish them from cluster members and the field population is to use
their chemistry. 

The derived abundances of Fe, $\alpha$-elements and Ba for the 5 stars of this
sample with
sufficient S/N are listed in Table 2. In Fig. \ref{alpha} they are compared with the 
sample of field giants
analysed by Gratton et al. (2003) and Venn et al. (2004), Terzan 8 stars by Carretta et al. (2014) and
Sagittarius stars by Carretta et al. (2010) and Sbordone et al. (2007).
In spite of the large uncertainties on the derived abundances it is clear that
the 5 stars have a metallicity which is not consistent with that of Terzan 8:
they span a wide range $-1.44<[Fe/H]<-0.83$ being significantly more metal-rich
than the cluster members (at [Fe/H]$\sim$-2.3).
%Only one star (\#94) has a low
%metallicity and an $\alpha$-enhancment compatible with the cluster. 
The comparison with the field
Milky Way and Sagittarius stars is less clear and is hampered by the large
uncertainties in the derived abundances. However, we note that while the
abundances of Si and Ca are consistent with those of both groups, the mean Ti I
and Mg abundance are significantly lower than those of the Galactic giants while
still compatible with those of Sagittarius\footnote{The Mg abundances of
three stars (\#47, \#57, \#65) lie $\sim0.4-0.6~dex$ ($\sim2~\sigma$) below the mean locus of Sgr
stars. Given the large involved uncertainties it is not clear whether this difference
is real or due to measurement errors.}. Moreover, Ba appears to be
enhanced by $\sim 0.3~dex$ in 4 of them, in agreement with what observed in Sgr
stars (Sbordone et al. 2007). Even more important is the excellent
agreement between the abundance of Fe measured from the neutral and ionized
lines in all the analysed stars. As stated in Sect. \ref{obs_sec} we adopted a gravity
derived assuming these objects as RGB stars at the same distance of Terzan 8.
Were these stars foreground Milky Way dwarfs their gravity would be larger
by $\Delta log~g>1$. According to the sensibility reported in Carretta et al.
(2014) this would produce a striking mismatch between the abundances of Fe I and
Fe II as a consequence of the wrong adopted ionization equilibrium. On the basis
of the above considerations we conclude that at least 4 out these 7 stars are
likely member of the Sagittarius stream.

\section{Conclusions}

We performed a detailed study of the structure and kinematics of the GC Terzan 8 by means of
a set of accurate radial velocities obtained through high-resolution spectroscopy
in the inner part of the cluster. 

The cluster density profile is well fitted by very extended models 
 with the largest half-mass radius among Galactic GCs ($22<r_{h}/pc<44$). 
By considering the estimated mass and the
Galactocentric distance of this GC we estimate an half mass-to-Jacobi radii
ratio of $r_{h}/r_{J}\sim0.16$.
In comparison with the other Galactic GCs it situates in the upper branch of the
$r_{h}/r_{J}~vs.~R_{GC}$ diagram constituted by extended low-mass GCs (Baumgardt
et al. 2010). The position of these clusters has been interpreted as a 
consequence of different cloud pressure and background tidal force at their
formation (Elmegreen 2008). It is interesting to note that many of these
clusters have been associated to merging dwarf galaxies. It is not clear if
Terzan 8 was born with such an extended structure or this is due to its dynamical 
evolution, although its long relaxation time suggests the former hypothesis.
In this context, if Terzan 8
spent most of its evolution within a dwarf galaxy (like the Sagittarius dSph) it
experienced a mild tidal field and could therefore form and evolve with a more 
extended 
structure with respect to other GCs formed within the Milky Way halo. This
could explain also the small fraction of Na-enhanced stars found in this
cluster (see Carretta et al. 2014). These stars are indeed expected to form
in the central part of the cluster after some 
$10^{6-7} yrs$ after the cluster formation being more resistant than the first
stellar generation to the tidal stripping process (D'Ercole et al. 2008;
Decressin et al. 2007). For
this reason, while in GCs the first stellar generation generally constitute only 
a small fraction of the cluster content, in Terzan 8 it is almost entirely 
retained in Terzan 8 because of the mild tidal stress experienced by this 
cluster.

We found no significant radial component of rotation within the
cluster, although slow rotations (with $v_{rot}~sin i/\sigma_{v}<0.4$) would not be
detected by our study and cannot therefore be excluded.

The derived overall velocity dispersion appears to be larger than what predicted
by dynamical models assuming kinetic energy equipartition and a moderate
amount of binaries. On the other hand, models without equipartition can
reproduce the observed velocity dispersion regardless of the adopted fraction of
binaries. The half-mass relaxation times of the considered models are
$1.7<t_{rh}/Gyr<7.5$, shorter than the estimated cluster age ($t_{9}$=12.2 Gyr;
Mar{\'{\i}}n-Franch et al. 2009). It is therefore unlikely that collisional effects
have not altered the kinematics of this GC. 
By means of Nbody simulations we estimated only a negligible contribution of 
tidal effects to the cluster velocity dispersion, at least when quasi-circular
orbits are considered. In principle, highly eccentric orbits could produce a more
significant inflation of the velocity dispersion even in the innermost regions
of the clusters. Such a hypothesis requires a specific study with extensive 
Nbody simulations. 
The observed discrepancy could be
eliminated if a large fraction ($\xi>30\%$) of binaries is present within Terzan
8. Unfortunately, the time coverage of our spectra does
not allow to identify binaries from their velocity variation. An estimate of the
binary fraction in this cluster has been provided by Milone et al. (2012) who
found $\xi=13.4\%$ within the ACS field of view. It is worth
noting that the uncertainties in the adopted distribution of periods,
eccentricity and mass-ratios of binaries affect the determination of the binary 
fraction through photometric and spectroscopic studies in different ways and can 
produce a mismatch between these two estimates.
It is also interesting to note that the masses of the models have been calculated from
the number counts on the CMD and can be therefore considered as an estimate of
the luminous mass of this cluster. The large observed velocity dispersion could
be viewed as an excess of dynamical mass within Terzan 8, as indicated by the
large dynamical mass-to-light ratios $(M/L_{V})_{dyn}>4$.
A systematic difference between the luminous and
dynamical masses of many GCs have been already noticed by Sollima et al. (2012)
who addressed the observed difference to the large uncertainties on the MF and
the retention fraction of dark remnants. Also in this case, different
assumptions on the dark remnants could alleviate or even eliminate the
difference between the observed and predicted velocity dispersion of this GC.

We detected a sample of 7 stars with radial velocities in the range
$100<v_{r}<200$ km/s apparently clustered around the systemic velocity of Terzan
8 but with a spread significantly larger than the cluster velocity dispersion.
These stars are spatially segregated with respect to the other field and cluster
stars occupying the same half of the observed field of view. The probability
that this occurs by chance is $\sim$0.8\%.
The abundance analysis performed on these stars indicates that they have a
metallicity which is inconsistent with that of the cluster. The abundance
patterns of the $\alpha$-elements and Ba indicate a deficiency of Mg and Ti I and an
overabundance of Ba in these
stars with respect to Galactic field stars and more similar to those measured in
Sagittarius stars. Moreover, the gravities estimated for these stars are
consistent with a distance similar to that of Terzan 8. The comparison with the
Galactic model of Robin et al. (2003) indicates that it is unlikely that
a significant number of Galactic field stars could be present at such a large
Galactocentric distance. We therefore conclude that they are likely member of
the Sagittarius stream surrounding Terzan 8. The presence of the Sagittarius 
stream has been already detected in some previous studies based on
wide field photometric surveys (Giuffrida et al. 2010; Siegel et al. 2011; 
Salinas et al. 2012). This however represents the first spectroscopic detection
of Sagittarius stars around this GC, confirming its belonging to the stream of
this satellite galaxy.

\section*{Acknowledgments}

This research has been founded by PRIN INAF 2011 "Multiple populations in globular
clusters: their role in the Galaxy assembly" (PI E. Carretta) and PRIN MIUR 
2010-2011 "The Chemical and Dynamical Evolution of the Milky Way and Local Group Galaxies" (PI F. Matteucci).
VD is an ARC Super Science Fellow. We thank the anonymous referee for
his/her helpful comments and suggestions.

\label{lastpage}

\end{document}